\documentclass{article}

\usepackage{amsmath}
\usepackage{graphicx}
\usepackage[round]{natbib}

\usepackage{hyperref}
\hypersetup{colorlinks=true, linkcolor=blue, filecolor=magenta, urlcolor=blue, citecolor=blue, pdfauthor=Zachary P. Neal}
\newcommand{\doi}[1]{\href{http://dx.doi.org/#1}{https://doi.org/#1}}

\usepackage{lastpage}

\begin{document}
\title{When can networks be inferred from observed groups?}
\author{Zachary P. Neal\\Michigan State University (zpneal@msu.edu)}
\date{}

\maketitle
\begin{abstract}
\noindent Collecting network data directly from network members can be challenging. One alternative involves inferring a network from observed groups, for example, inferring a network of scientific collaboration from researchers' observed paper authorships. In this paper, I explore when an unobserved undirected network of interest can accurately be inferred from observed groups. The analysis uses simulations to experimentally manipulate the structure of the unobserved network to be inferred, the number of groups observed, the extent to which the observed groups correspond to cliques in the unobserved network, and the method used to draw inferences. I find that when a small number of groups are observed, an unobserved network can be accurately inferred using a simple unweighted two-mode projection, provided that each group's membership closely corresponds to a clique in the unobserved network. In contrast, when a large number of groups are observed, an unobserved network can be accurately inferred using a statistical backbone extraction model, even if the groups' memberships are mostly random. These findings offer guidance for researchers seeking to indirectly measure a network of interest using observations of groups.\\\\
\textbf{NOTE: This is a post-print. The version of record can be found at: Neal, Z. P. (2024). When can networks be inferred from observed groups? \textit{Network Science, 12}, 189-200. \url{https://doi.org/10.1017/nws.2024.6}}
\end{abstract}

\maketitle

\section{Introduction}
Collecting network data directly from network members, for example through surveys or interviews, can be challenging due to the resource intensive nature of the data collection and the risks to data quality from missingness, reporting errors, and reactivity \citep{adams2020gathering,marsden2011survey}. These challenges have led network researchers to seek alternate, indirect measurement methods. When the network of interest is undirected, one common alternative involves inferring an unobserved network from observed groups such as club memberships or event participations, using a suitably-binarized projection \citep[e.g.,][]{breiger1974duality,newman2004coauthorship, mizruchi1996interlocks,andris2015rise,schaefer2010fundamental}. However, little is known about the circumstances under which a network inferred from observed groups accurately captures the unobserved network of interest. That is, \textit{when can networks be inferred from observed groups}?

To answer this question, I perform a series of experiments, varying the structure of the unobserved network being inferred, characteristics of the observed groups, and the method used to infer a network from the groups. When a small number of groups are observed, an unobserved network can be accurately inferred using a simple unweighted two-mode projection, provided that each group's membership closely corresponds to a clique in the unobserved network. In contrast, when a large number of groups are observed, an unobserved network can be accurately inferred using a statistical backbone extraction model, even if the groups' memberships are mostly random. These findings suggest that networks can be inferred from observed groups, and offers guidance on when such inferences are sufficiently accurate to be used when data cannot be collected directly from network members.

The remainder of the paper is organized in four sections. In the background section, I review the potential of inferences from observed groups as a possible solution to challenges to directly collecting network data. In the methods section, I describe an experiment designed to evaluate the accuracy of a network inferred from observed groups. In the results section, I report the accuracy of networks inferred under experimentally-varied conditions, highlighting when such inferences are and are not accurate. Finally, in the discussion section, I identify opportunities for future research and offer recommendations for researchers wishing to infer networks from observed groups.

\section{Background}
One common approach to collecting network data is to collect data directly from the network's members. For example, if we want to know who your friends are, there is a strong intuitive appeal to simply asking you ``Who are your friends?'' Although there are many variations, direct collection of network data typically takes place via a survey or interview, which includes one or more `name generator' questions like the one above \citep{adams2020gathering,marsden2011survey}. However, the direct collection of network data comes with a number of challenges: it can be resource-intensive \citep{adams2020gathering,marsden2011survey}, it is subject to measurement error \citep{wang2012measurement} and missingness \citep{kossinets2006effects}, and it may be impossible when network members are too young \citep{neal2020systematic} or not human \citep{krause2009animal}.

The severity of these challenges varies by context, and strategies exist for overcoming them. However, these challenges have led network researchers to look for indirect methods of collecting network data. Among the most widely used approaches involves inferring an unobserved network from observed groups \citep[e.g.,][]{newman2004coauthorship,mizruchi1996interlocks,andris2015rise,schaefer2010fundamental}. In this paper, I define a `group' simply as a collection of individuals whose structure is unspecified (e.g., a party's list of attendees, but not who talked to whom), and a `network' as a structure among individuals \citep[e.g., who talks to whom;][]{wellman1988structural,neal2023duality}. A key advantage to this approach over direct data collection is that group ``affiliations are often observable from a distance (e.g., government records, newspaper reports), without having to have special access to the actors'' \citep{borgatti2011analyzing}. I focus on contexts where the number of observed or observable groups $G$ is at least as large as the number of $N$ actors who might affiliate with those groups (i.e., where $G \geq N$). This often occurs in contexts where membership in many groups can be discerned from archival data, or collected through field-based observations conducted over an extended period.

\cite{breiger1974duality} provided the most well-known illustration of how a one-mode network could be derived from two-mode data about individuals' group affiliations, inferring a network among 18 women from observations of their attendance at 14 social events. This approach proposes that a network of shared group affiliations (i.e., a bipartite or two-mode projection) provides some information about the network connections among the groups' members. It relies on the logic that if two people belong to many of the same groups or participate in many of the same events (what \cite{feld1981focused} called `foci'), then they likely interact and have or will form ties. 

Transforming two-mode data into one-mode data via projection necessarily involves the loss of some information. Nonetheless, it has been used to indirectly measure networks in a wide range of contexts, and in some fields has become the \textit{de facto} standard approach. Unobserved networks of scientific collaboration are inferred from researchers' observed paper authorships \citep[e.g.,][]{newman2004coauthorship}, unobserved networks of corporate executives are inferred from their observed board memberships \citep[e.g.,][]{mizruchi1996interlocks}, unobserved networks of political alliance are inferred from lawmakers observed memberships in voting blocs \citep[e.g.,][]{andris2015rise}, and unobserved social networks are inferred from young childrens' play groups \citep[e.g.,][]{schaefer2010fundamental}. However, despite its widespread use, it remains unknown whether or when a network inferred from observed groups is accurate.

\begin{figure}
\centering
\includegraphics[width=\columnwidth]{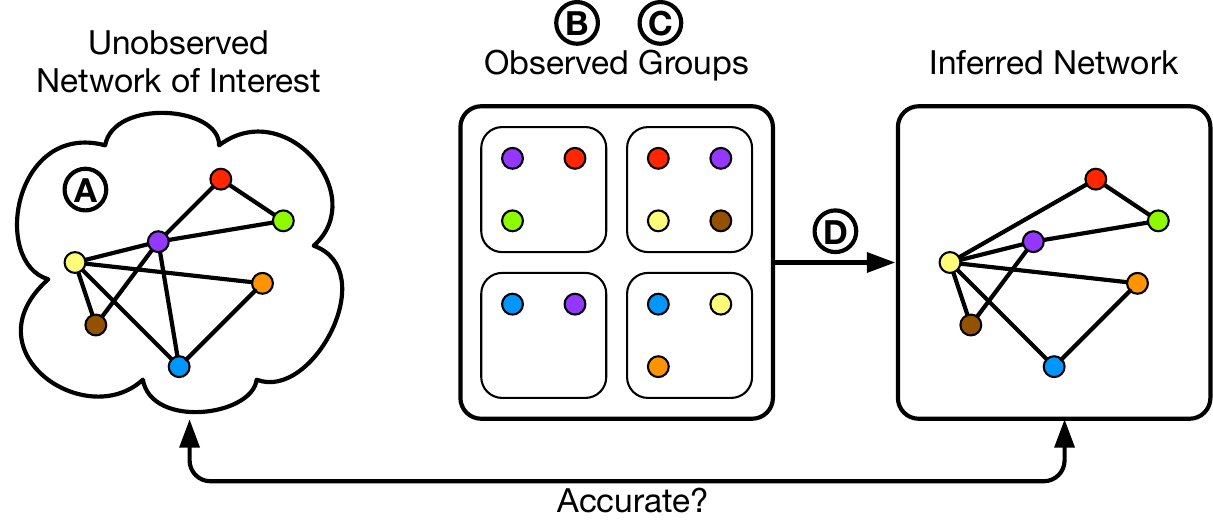}
\vspace*{10mm}
\caption{Relationship between an unobserved network, observed groups, and inferred network. Accuracy may depend on (A) the structure of the unobserved network, (B) the number of observed groups, (C) the extent to which observed group correspond to cliques in the unobserved network, and (D) the method used to infer network relationships from group memberships.}
\label{fig:framework}
\end{figure}

But, what does it mean to accurately infer a network from observed groups? Figure \ref{fig:framework} illustrates the relationship between an unobserved network, observed groups, and inferred network \citep{peel2022statistical}. On the left is a one-mode network of interest depicting the connections between seven \textit{agents} (e.g., people). Although this network exists, we can not directly observe it, perhaps because these agents declined to complete a network survey. Instead, we can only observe these agents' memberships in \textit{groups} (e.g. attending events together, belonging to the same club, etc.), which are driven at least in part by their unobserved network ties \citep{feld1981focused,schaefer2022youth,neal2023duality}. The example in Figure \ref{fig:framework} illustrates four different observed groups. These observed groups are simply sets of agents observed together (e.g., the top left group includes the purple, red, and green agents), but do not contain any information about the structure among their members. Notably, in some cases, group membership corresponds to a clique in the unobserved network (e.g., the top left group), while in other cases it does not (e.g., the top right group). These observed groups can be summarized in a two-mode network in which agents are connected to groups, and via projection, we can transform this two-mode agents-to-groups network into a one-mode agents-to-agents network using a projection. Accuracy in this context refers to the extent to which the one-mode network obtained via this process is similar to the unobserved network of interest. The goal, as \cite{peel2022statistical} explain, is to accurately infer or `reconstruct' a network of interest from indirect data, here of observed groups. In the \textit{supplementary materials} I explore the related goal of inferring key characteristics of the network of interest, without actually reconstructing the network itself.

At least four factors might influence how accurately an unobserved network can be inferred from observed groups: (A) the structure of the unobserved network, (B) the number of observed groups, (C) the extent to which observed groups correspond to cliques in the unobserved network, and (D) the method used to infer network relationships from group memberships. First, networks with certain structures may be easier to accurately infer than others. For example, because the projection of any bipartite network, ``even a random bipartite network...will be highly clustered'' \citep[][p. 128]{watts2003}, it may be easier to accurately infer an unobserved one-mode network whose structure is clustered. Second, inferences may be more accurate when they are based on a large number of observed groups because such inferences can draw on more information. Third, inferences may be more accurate when they are based on observed groups whose membership closely corresponds to cliques in the unobserved network \citep{guillaume2004bipartite}. Finally, the accuracy of inferences may depend on how network edges are inferred from observed group memberships.

\section{Methods}
To understand when networks can be inferred from observed groups, I perform a series of experiments that follow Figure \ref{fig:framework}. First, I generate a one-mode network using one of five network models, or choose one of five one-mode empirical networks, that serves as the hypothetical unobserved network of interest (see \ref{sec:networks}). Second, I simulate the groups that a researcher might observe as a result of this network, varying both their number and correspondence to cliques in the unobserved network (see \ref{sec:groups}). Third, I infer a one-mode network from these observed groups, using either a simple unweighted two-mode projection or a backbone extracted using the stochastic degree sequence model (see \ref{sec:inference}). Finally, I compute the similarity of the unobserved network of interest and the inferred network (see \ref{sec:analysis}).

\subsection{Unobserved networks}\label{sec:networks}
When attempting to infer a unobserved network from observed groups, the structure of the unobserved network is unknown, but may nonetheless impact the accuracy of the inferences. Therefore, I explore the accuracy of inferences when the unobserved network has a range of structures, using both artificial and empirical networks.

Artificial networks generated using well-known network models are useful because they have well-known structural properties. Here, I consider five such models. First, I generate 50-node \textit{random networks} using the Erd\H{o}s-R\'enyi model, where the probability of an edge is 0.08 \citep{erdosrenyi}. Second, I generate 50-node \textit{small-world networks} using the Watts-Strogatz model, where each node in a ring lattice is initially connected to its four nearest neighbors, then edges are re-wired with probability 0.05 \citep{watts1998collective}. Third, I generate 50-node \textit{scale-free networks} using the preferential attachment model, where two edges are added in each step \citep{barabasi1999emergence}. Fourth, I generate 50-node \textit{caveman} networks that contain 10 cliques of 5-nodes each \citep{watts1999networks}. Finally, I generate 50-node \textit{core-periphery} networks in which 10 nodes form a dense core ($d = 0.85$), and 40 nodes in the periphery are connected to 1 or 2 core nodes \citep{borgatti2000models}. These specifications all yield networks containing 50 nodes and about 100 edges, and therefore hold network size and density constant.

Although these artificial network models are well understood, they can generate networks that may not resemble real networks. Therefore, I also consider five empirical networks: the interactions of 62 Dolphins in Doubtful Sound \citep{lusseau2003bottlenose}, the marital network among 15 families in 15\textsuperscript{th} century Florence \citep{padgett1993robust}, the social relationships among 34 members of a Karate club \citep{zachary1977information}, the friendships among 71 lawyers in a Northeastern US corporate law firm \citep{lazega2001collegial}, and the friendships among 32 workers in a tailor shop in Zambia \citep{kapferer1972strategy}. Although these empirical networks vary slightly in size and density, they are sufficiently similar to the five artificial networks to permit comparisons.

\subsection{Observed groups}\label{sec:groups}
When attempting to infer a unobserved network from observed groups, characteristics of the observed groups may impact the accuracy of the inferences. If every observed group corresponds to a clique in the unobserved network, and a sufficiently large number of groups are observed, then the unobserved network can be inferred with perfect accuracy using a two-mode projection of the observed groups \citep{guillaume2004bipartite}. This is closely related to the NP-hard `clique cover problem,' which involves finding the smallest number of cliques that completely cover a network \citep{karp1972}. Although an unobserved network can be accurately inferred under these conditions, in practice a researcher may only be able to observe a limited number of groups, or may only observe groups that do not perfectly correspond to cliques in the unobserved network. Therefore, I experimentally vary both the number of observed groups, and the extent to which group memberships correspond to cliques in the unobserved network.

First, given an unobserved network of $N$ nodes, I consider the accuracy of a network inferred from observations of $N$, $2N$, $5N$, $10N$, $20N$, or $50N$ groups. Inferences drawn from more observed groups should be more accurate because they are based on more information. The lower end of this experimental range (i.e., observing $N$ groups) represents the fewest number of observed groups from which any network structure could, in principle, be inferred \citep{neal2012structural}. The upper end of this experimental range (i.e., observing $50N$ groups) represents the largest number of groups that might typically be observable. For example, \cite{neal2020sign} inferred a network among 100 U.S. Senators from an average of 3500 bill sponsorships (i.e., $35N$ groups).

Second, I consider the accuracy of a network inferred from observed groups whose members have a 50\%, 60\%, 70\%, 80\%, 90\%, or 100\% chance of being members of the same clique in the unobserved network. Inferences drawn from observed groups that more closely correspond to cliques in the unobserved network should be more accurate because they contain more information, and less noise, about the structure of the network. The upper end of this experimental range (i.e., 100\%) represents a scenario in which each observed group's membership is simply a clique in the unobserved network. For example, a tightly-knit clique of friends may be observed hanging out (i.e. an observed group) with no one else present. The lower end of this experimental range (i.e., 50\%) represents a scenario in which members of observed groups may or may not be members of the same clique in the unobserved network. For example, a group of researchers may be observed writing a grant together (i.e., an observed group), but only some of them are collaborators (i.e., they are not a clique). In the \textit{supplementary materials} I also consider one case outside these experimental conditions, where a very large number of groups are observed ($200N - 1000N$), but members of those groups are highly unlikely to be members of a clique ($p = 0.1$).

I use a model of team formation \citep{guimera2005team} that has been formalized as a two-mode generative model \citep{neal2023duality} to simulate the memberships of groups that might be observed. The model first randomly chooses a clique from the unobserved network. Given a clique containing $k$ nodes, it then generates an observed group of $k$ members by filling each position with either a member of the clique (with probability $p$) or someone else (with probability $1-p$). When $p = 1$, the observed group's members are simply the clique's members. In contrast, when $p = 0.5$, the observed group's members may or may not be the clique's members.

This approach involves the analysis of simulated groups that a researcher might observe, as opposed to actual groups that a researcher did observe. However, this generative model has previously been shown to generate simulates group that have characteristics of empirically-observed groups \citep{neal2023duality}. Additionally, using a generative model offers an important advantage over using empirical data: it is possible to experimentally manipulate how many groups are observed, and the extent to which those groups correspond to cliques, and therefore to investigate the hypothesized role that these two factors play in the accuracy of inferred networks.

\subsection{Inferring a network}\label{sec:inference}
Given a set of observed groups organized as a two-mode network, a weighted one-mode network can be derived via projection, where the edge weights indicate the number of times two nodes were observed in the same group. There are many ways to handle these edge weights when the goal is to infer an unweighted one-mode network \citep{borgatti2011analyzing}. In this experiment, I compare the accuracy of inferences drawn using a simple approach to those drawn using a state-of-the-art statistical backbone extraction model.

The simplest and most widely-used approach for handling edge weights in a projection is to ignore them, and to focus on a simple unweighted projection. In an unweighted projection, two nodes are connected if they were observed in \textit{one or more} of the same groups. This approach offers simplicity and computational efficiency, but sets a low threshold for inferring that two nodes are connected in the unobserved network of interest. Other, higher thresholds can be used (e.g., observed in two, three, or more of the same groups), but the choice of a given threshold is arbitrary. This approach also ensures that the inferred network will be dense, with high levels of transitivity and clustering, regardless of the true structure of the unobserved network \citep{latapy2008basic,neal2014backbone,watts2003}, which may diminish its accuracy.

Although methods have been proposed for choosing an edge weight threshold or normalizing edges weights in a projection \citep{borgatti2011analyzing}, the current state-of-the-art for obtaining an unweighted projection are statistical backbone extraction models. These models use information from the two-mode data (here, the observed groups) to define a statistical null model, then test the statistical significance of each edge's weight to determine which should be retained in an unweighted `backbone.' Many backbone extraction models exist, however only two have preliminary evidence that they can accurately infer unobserved networks: the stochastic degree sequence model (SDSM) and fixed degree sequence model \citep[FDSM;][]{neal2021comparing,neal2022inferring,gomes2022network}. Although both SDSM and FDSM are candidates for inferring an unobserved network from observed groups, I consider only the former because prior work has demonstrates they yield similar results \citep{neal2021comparing} and because FDSM is too computationally intensive to be useful in practice \citep{godard2022fastball}.

The formal specification of the SDSM is described by \cite{neal2021comparing}, but like all statistical backbone extraction models it aims to determine when an edge weight in a projection is statistically significantly larger than the weight that would be expected in the projection of a random two-mode network. The SDSM is distinguished from other backbone models by the information from the two-mode network that it uses to evaluate the significance of an edge weight. Specifically, it evaluates whether a given edge's weight in a projection is larger than expected in a random null model that simultaneously controls for the degree sequences of both types of nodes. In this context, it evaluates whether the number of group memberships shared by two individuals (i.e. the edge weight in a projection) is larger than expected in a null model that simultaneously controls for (a) the number of groups to which each of those individuals belong and (b) the number of individuals that belong to each group. By considering this information, the SDSM applies a unique threshold to each edge. For example, observing two people in many of the same \textit{small} groups such as dinner parties provides stronger evidence for inferring they are connected than observing them in many of the same \textit{large} groups such as concerts. 

\subsection{Experimental design and analysis}\label{sec:analysis}
Table \ref{tab:design} summarizes the factorial experimental design, which varies 10 unobserved network structures, 6 numbers of observed groups, 6 probabilities that groups correspond to cliques, and 2 inference methods, for a total of 720 experimental conditions. Within each condition, I compute the accuracy of the inferred network as the similarity between the unobserved network and the inferred network, averaged over 1000 replications. There are several ways to measure the similarity of two networks. In the results below I report the Pearson correlation coefficient, which is also known as the Matthews correlation coefficient in the context of evaluating binary classifications (i.e., is the edge present or absent?), because it is more robust than alternate metrics \citep{chicco2020advantages}. However, sensitivity analyses confirm that other metrics, including Cohen's $\kappa$ and the Jaccard coefficient, yield the same results. The supplementary materials and the code necessary to replicate all results reported below is available at \url{https://osf.io/6vcxa}.

\begin{table}[]
\begin{tabular}{ll}
\hline
Factor & Levels \\
\hline
(A) Unobserved network & Random, Small World, Scale Free, Caveman, Core-Periphery \\
& Dolphin, Florentine, Karate, Law, Tailor \\
(B) Number of observed groups & $N$, $2N$, $5N$, $10N$, $20N$, or $50N$ \\
(C) Groups are cliques? & 50\%, 60\%, 70\%, 80\%, 90\%, or 100\% \\
(D) Method & Unweighted projection \\
& Stochastic degree sequence model (SDSM) \\
 & \\
Replications per condition & 1000 \\
Outcome & Mean Correlation, Jaccard, Kappa \\
\hline
\end{tabular}
\vspace*{2mm}
\caption{Summary of factorial experimental design}
\label{tab:design}
\end{table}

\section{Results}
\begin{figure}
\centering
\includegraphics[width=\columnwidth]{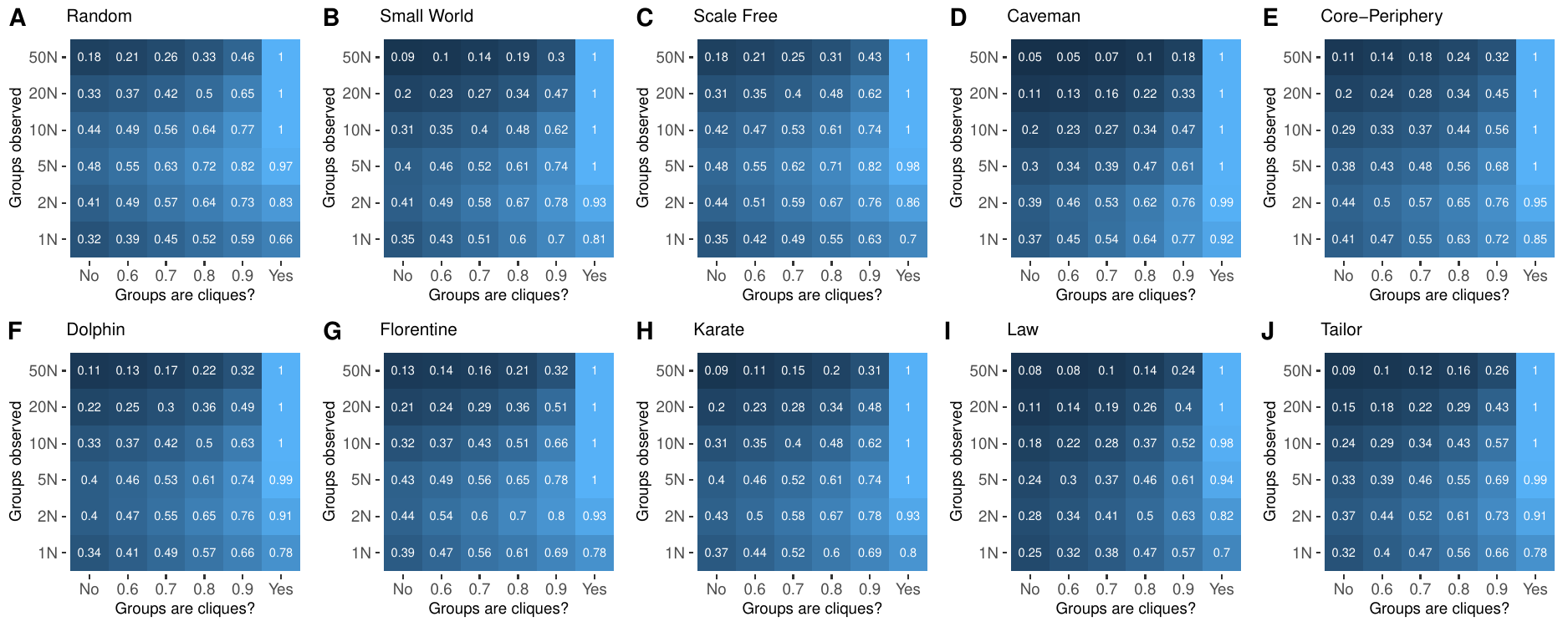}
\vspace*{10mm}
\caption{Accuracy of a network inferred from observed groups using an unweighted projection, by (a) the structure of the unobserved network being inferred, (b) number of groups observed, and (c) extent to which the observed groups correspond to cliques. Accuracy is measured using the correlation between the unobserved and inferred networks.}
\label{fig:naive}
\end{figure}

\begin{figure}
\centering
\includegraphics[width=\columnwidth]{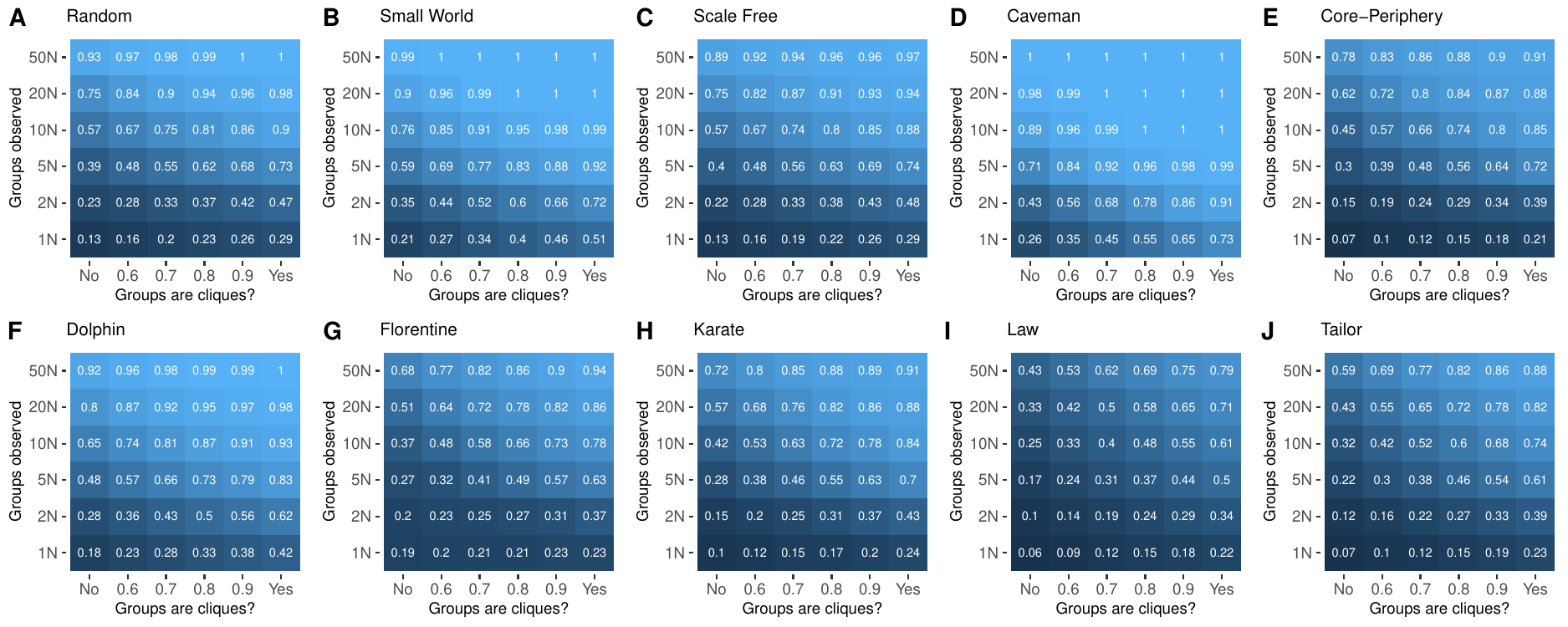}
\vspace*{10mm}
\caption{Accuracy of a network inferred from observed groups using a backbone extracted with the stochastic degree sequence model, by (a) the structure of the unobserved network being inferred, (b) number of groups observed, and (c) extent to which the observed groups correspond to cliques. Accuracy is measured using the correlation between the unobserved and inferred networks.}
\label{fig:sdsm}
\end{figure}

Figure \ref{fig:naive} reports the accuracy of a network inferred from observed groups using an unweighted projection, by the structure of the unobserved network being inferred (panels), number of groups observed (y-axis within panels), and extent to which the observed groups correspond to cliques (x-axis within panels). Similarly, Figure \ref{fig:sdsm} reports the accuracy of a network inferred from observed groups using a backbone extracted with the stochastic degree sequence model. In both cases, the accuracy of the inferred network is measured using the mean correlation between the unobserved `true' network and the inferred network over 1000 replications for the given experimental condition. Lighter shades represent higher correlations, and thus conditions under which inferences are more accurate.

Table \ref{tab:results} summarizes the experimental outcomes illustrated in Figures \ref{fig:naive} and \ref{fig:sdsm} via regression by predicting the accuracy of an inference as a function of the unobserved network's topology (size, density, transitivity, and number of cliques) and characteristics of the observed groups (mean number of group members per group, mean number of group memberships per person, number of observed groups, and extent to which observed groups are cliques). Unstandardized ($B$) and standardized ($\beta$) estimates are reported. Standard errors and $p$-values are not reported because, in a simulation context where sample size is arbitrary, they are not meaningful. When interpreting these results below, I focus on the standardized estimates because they indicate which properties have relatively more or less impact on the accuracy of inferences.

\begin{table}[]
\begin{tabular}{lcc|cc}
\hline
 & \multicolumn{2}{c}{\underline{Unweighted Projection}} & \multicolumn{2}{c}{\underline{SDSM Backbone}} \\
 & $B$ & $\beta$ & $B$ & $\beta$ \\
\hline
Intercept & -0.143 & --- & 0.117 & --- \\
Size & -0.001 & -0.034 & 0.004 & 0.224 \\
Density & -0.150 & -0.041 & 0.234 & 0.060 \\
Transitivity & -0.140 & -0.087 & 0.789 & 0.452 \\
Number of Cliques & $<$ 0.001 & 0.191 & $<$ 0.001 & 0.211 \\
Mean Group Members & -0.017 & -0.064 & -0.177 & -0.611 \\
Mean Group Memberships & -0.001 & -0.256 & -0.003 & -0.499 \\
Groups observed & -0.004 & -0.287 & 0.016 & 0.922 \\
Groups are cliques? & 1.17 & 0.76 & 0.511 & 0.304 \\
$R^2$ & \multicolumn{2}{c}{0.749} & \multicolumn{2}{c}{0.677} \\
\hline
\end{tabular}
\vspace*{4mm}
\caption{Regression predicting the accuracy of an network inferred using a given approach, as a function of the unobserved network's topology and characteristics of the observed groups.}
\label{tab:results}
\end{table}

Turning first to inferences drawn using a simple unweighted projection, several patterns appear in Figure \ref{fig:naive}. First, as described by \cite{guillaume2004bipartite}, when a large number of groups are observed, and those groups directly correspond to cliques in the unobserved network, the unobserved network can be inferred with perfect accuracy ($r = 1$). Second, as expected, inference accuracy declines when the observed groups are less likely to correspond to cliques in the unobserved network. Third, unexpectedly, inferences are \textit{not} more accurate when they are based on a larger number of observed groups. Instead, inferences are most accurate when they are based on $2$-$5N$ observed groups. Inferences based on fewer observed groups are less accurate because they draw on less information, while inferences based on more observed groups are less accurate because they are overwhelmed by noisy information. Finally, these patterns are consistent across all ten types of unobserved network, suggesting that the structure of the unobserved network plays little role in the accuracy of inferences drawn using an unweighted projection.

These patterns are confirmed in the left panel of Table \ref{tab:results}. The most important factor in accurately inferring an unobserved network from observed groups using an unweighted projection is the extent to which the observed groups correspond to cliques ($\beta = 0.76$). The more closely the observed groups correspond to cliques, the more accurately a network can be inferred using an unweighted projection. All other characteristics of the unobserved network and observed groups have a limited impact on the accuracy of inferences.

Turning next to inferences drawn using a backbone extracted with the stochastic degree sequence model, several patterns also appear in Figure \ref{fig:sdsm}. First, as expected, the accuracy of inferences is higher when more groups are observed. This occurs because, as with any statistical inference model, inferences are more accurate when they are based on more data (here, when they are based on more observed groups). Second, also as expected, the accuracy of inferences is higher when the observed groups more closely correspond to cliques in the unobserved network. Finally, there is some variation in the accuracy of inferences for different unobserved networks. For example, across all experimental conditions, an unobserved caveman network can be inferred with high accuracy (mean $r = 0.84$), while the Law Firm network can be inferred with much lower accuracy (mean $r = 0.38$).

These patterns are confirmed in the right panel of Table \ref{tab:results}. The most important factor in accurately inferring an unobserved network from observed groups using an SDSM backbone is the number of groups observed ($\beta = 0.922$). The more groups that are observed, the more accurately a network can be inferred using an SDSM backbone. Other factors play a more limited role. For example, unobserved networks with higher transitivity can be inferred more accurately ($\beta = 0.452$), which helps explain the generally higher accuracy with which a caveman network can be inferred. Similarly, inferences are more accurate when the setting is characterized by smaller groups ($\beta = -0.611$) and individuals with fewer group memberships ($\beta = -0.499$).

Comparing the accuracy of inferences drawn using these two approaches suggests that an SDSM backbone yields slightly more accurate inferences (mean $r = 0.61$) than an unweighted projection (mean $r = 0.51$). However, there are significant variations that impact when each approach is likely to yield an accurate inference. The estimates in Table \ref{tab:results} indicate that when inferences are based in more observed groups, the accuracy of an unweighted projection is reduced, while the accuracy of an SDSM backbone is increased. Likewise, although the extent to which groups correspond to clique has a large impact on the accuracy of an unweighted projection, it plays a less significant role in the accuracy of an SDSM backbone. The case of inferring an unobserved 50-node random network serves to illustrate these differences. If 250 groups are observed (i.e., $5N$) and those groups directly correspond to cliques in the unobserved network (i.e., $p = 1$), then an unweighted projection offers a very accurate representation of the unobserved network ($r = 0.97$), while an SDSM backbone is less accurate ($r = 0.73$). In contrast, if 2500 groups are observed (i.e., $50N$) but those groups do not correspond to cliques in the unobserved network ($p = 0.5$), then an SDSM backbone offers a very accuracy representation of the unobserved network ($r = 0.93$), while an unweighted projection is much less accurate ($r = 0.18$).

\section{Discussion}
Practical and methodological challenges associated with collecting network data directly from network members have led network researchers to develop indirect data collection methods. Among the most widely used methods involves attempting to infer an unobserved network from observed groups, for example, inferring an unobserved network of collaboration from observed participation on published papers. Although this approach is widely used, little is known about when networks can be accurately inferred from observed groups.

In this paper, I conducted a series of experiments to examine how the accuracy of a network inferred from observed groups depends on four factors: the structure of the unobserved network to be inferred, the number of groups observed, the extent to which the observed groups correspond to cliques in the unobserved network, and the method used to draw inferences. The results demonstrate that on average an unobserved network can be inferred from group observations with moderate accuracy (mean $r = 0.55$), but that there is substantial variation in the expected accuracy of inferences under different circumstances (range $r = 1 - 0.05$). 

These findings provide researchers with guidance about when an unobserved network can be accurately inferred from observed groups. First, researchers can use a simple unweighted projection to accurately (mean $r = 0.84$) infer an unobserved network of $N$ nodes if $2N$ to $5N$ groups are observed \textit{and} membership in those groups are believed to closely correspond to cliques in the unobserved network of interest ($p \geq 0.9$). However, the ability to accurately infer an unobserved network under these circumstances may not be especially useful in practice because often observed groups will not perfectly correspond to cliques in an unobserved network, and even if they did, it would be impossible to know.

Second, researchers can use an SDSM backbone to accurately (mean $r = 0.8$) infer an unobserved network of $N$ nodes if at least $10N$ groups are observed. Networks inferred using an SDSM backbone remain reasonably accurate even when the membership of observed groups do not correspond to cliques in the unobserved network (when $p \leq 0.6$, mean $r = 0.7$). The ability to accurately infer an unobserved network under these circumstances is useful in practice because archival data sources mean the number of observed groups is often much larger than the number of nodes \citep{borgatti2011analyzing}, and because the relationship of groups to cliques is usually unknown.

Finally, researchers should not infer an unobserved network of $N$ nodes based on $N$ or fewer observed groups because such inferences will be inaccurate (mean $r = 0.4$). This imposes an important scope condition on inferring networks from observed groups, and limits the applicability of this approach in some contexts. For example, despite serving as an early example \citep{breiger1974duality}, it may be difficult to accurately infer a social network among 18 women from just 14 social events. Similar issues arise in more contemporary multi-level networks, where the number of scientists (nodes) exceeds the number of disciplines \citep[groups;][]{bellotti2016comparing}, the number of managers (nodes) exceeds the number of organizations \citep[groups;][]{brennecke2016knowledge}, or the number of students (nodes) exceeds the number of extra-curricular activities \citep[groups;][]{schaefer2022youth}.

Although these results suggest that an unobserved network can be inferred from observed groups under certain circumstances, this indirect approach to measurement should be used with caution. When a network is measured directly by asking network members a name generate question (e.g., who are your \textit{friends}) or from archival data (e.g., who do you \textit{follow} online), the meaning of edges in the network are explicitly known (friendship or following). In contrast, when a network is inferred from observed groups, the meaning of edges in the inferred network is ambiguous and depends on why (and, indeed, whether) group co-membership suggests a relationship between two nodes. Consider a network inferred from observations of groups of legislators sponsoring bills, as is common in political network research. The edges in such a network might be interpreted as indicating relationships of strategic political alliances because co-sponsorship requires coordinated legislative action, or of communication because co-sponsorship requires talking to one another about bills, or merely of ideological similarity because co-sponsorship indicates that two legislators favor the same bills \citep{neal2022constructing}. Therefore, when a network is inferred from observed groups, the researcher must offer a theory or rationale that observations of shared group memberships provides a valid indicator of a particular type of relationship. Relatedly, the researcher must also offer a theory or rationale that shared group memberships observed over a given time period provides a valid indicator of a given cross-sectional network.

As a first exploration into when an unobserved network can be inferred from observed groups using projection-based methods, this study points to several directions for future research. First, while inference using an SDSM backbone seems promising, this method relies on frequentist $p$-values generated with reference to a null model conditioned on two only characteristics of observed groups: groups' sizes and individuals' number of memberships. Future research may explore developing new backbone models that may improve inferential accuracy by computing Bayesian likelihoods of edges' existence, or by using an ERGM framework to condition the null model on additional characteristics. Second, these results are based on simulated observations of independent groups. Future research may explore the accuracy of inferences from groups that have been empirically observed, and from groups whose membership is not independent.

A half-century ago, \cite{breiger1974duality} illustrated how a one-mode network could be constructed from information about observed groups organized as a two-mode network. This approach has since become widely used as a way to indirectly measure one-mode networks that would be impractical or impossible to measure directly. However, as an indirect measurement, it has been unclear whether networks inferred from observed groups in this way are accurate, that is, whether they correctly capture the structure of the unobserved network of interest. These experimental results indicate that they can, thereby vindicating the approach described by \cite{breiger1974duality} as a way to indirectly measure networks. However, they also demonstrate that the degree of accuracy depends on several factors, and therefore they also provide much-needed guidance on when such an approach is appropriate.

\section*{Data availability statement}
The supplementary materials and the code necessary to replicate all results reported below is available at \url{https://osf.io/6vcxa}.

\section*{Funding statement}
This work was supported by National Science Foundation awards \#2016320 and \#2211744.

\section*{Competing interests statement}
None.

\end{document}